\documentclass[12pt]{article}

\usepackage{cite}
\usepackage{mathptmx}
\usepackage{setspace}
\usepackage{multirow}
\usepackage{longtable}

\usepackage{changepage}
\usepackage{dcolumn}
\usepackage[table]{xcolor}
\usepackage{floatrow}
\usepackage{float}
\usepackage{subcaption}
\usepackage{graphicx,xcolor} 
\usepackage{url}
\usepackage{hyperref}
\usepackage{ulem}
\usepackage{authblk}

\usepackage[margin=1in]{geometry}
\usepackage{amsmath}
\usepackage{amssymb}
\usepackage{pgf}

\definecolor{DarkCyan}{rgb}{ 0 0.545 0.937}
\definecolor{LightCyan}{rgb}{ 0.28 0.82 0.8}
\definecolor{myblue-strong}{rgb}{0.4 0.67 0.84}
\definecolor{myblue-medium}{HTML}{A3CCE7}
\definecolor{myblue-weak}{HTML}{D1E5F3}
\definecolor{myblue-negligible}{HTML}{EFF6FB}

\definecolor{table-color-dark}{HTML}{E4D2BA}
\definecolor{table-color-light}{HTML}{F6F0E8}

\definecolor{myred-strong}{rgb}{0.91 0.60 0.46}
\definecolor{myred-medium}{HTML}{F1C0AB}
\definecolor{myred-weak}{HTML}{FAEAE3}
\definecolor{myred-negligible}{HTML}{FCF4F1}

\newcommand{\eref}[1]{Eq.~(\ref{#1})}%
\def\bea{\begin{eqnarray}}
\def\eea{\end{eqnarray}}

\newcolumntype{d}[1]{D..{#1}}

\title{The fate of the American dream: A first passage under resetting approach to income dynamics}
\author{Petar Jolakoski$^{1}$, Arnab Pal$^{2\footnote{ \href{mailto:arnabpal@imsc.res.in}{arnabpal@imsc.res.in} }}$, Trifce Sandev$^{1,3,4}$, Ljupco Kocarev$^{1,5}$,\\ Ralf Metzler$^{3,6}$, Viktor Stojkoski$^{7,8\footnote{Corresponding author: \href{mailto:viktor.stojkoski@univ-toulouse.fr}{viktor.stojkoski@univ-toulouse.fr} }}$}

\affil{%
\footnotesize
$^{1}$Research Center for Computer Science and Information Technologies, Macedonian Academy of Sciences and Arts, Bul. Krste Misirkov 2, 1000 Skopje, Macedonia \\
\footnotesize
$^{2}$The Institute of Mathematical Sciences, CIT Campus, Taramani, Chennai 600113, India \& 
Homi Bhabha National Institute, Training School Complex, Anushakti Nagar, Mumbai 400094,
India  \\
\footnotesize
$^{3}$Institute of Physics \& Astronomy, University of Potsdam, D-14776 Potsdam-Golm, Germany \\  \footnotesize
$^{4}$Institute of Physics, Faculty of Natural Sciences and Mathematics, Ss.~Cyril and Methodius University, Arhimedova~3, 1000 Skopje, Macedonia\\
$^{5}$Faculty of Computer Science and Engineering, Ss. Cyril and Methodius University, PO Box 393, 1000 Skopje, Macedonia \\
\footnotesize
\footnotesize $^{6}$ Asia Pacific Centre for Theoretical Physics, Pohang 37673, Republic of Korea\\
\footnotesize
$^{7}$Center for Collective Learning, ANITI, University of Toulouse \\
\footnotesize
$^{8}$Faculty of Economics, Ss.~Cyril and Methodius University}

\date{\today}

\begin{document}
\maketitle

\begin{abstract}
Detailed knowledge of individual income dynamics is crucial for investigating the existence of the American dream: Are we able to improve our income status during our working life? This key question simply boils down to observing individual status and how it moves between two thresholds: the current income and the desired income. Yet, our knowledge of these temporal properties of income remains limited since we rely on estimates coming from transition matrices which simplify income dynamics by \textit{aggregating the individual changes into quantiles} and thus overlooking significant microscopic variations. Here, we bridge this gap by employing \textit{First Passage Time} concepts in a baseline stochastic process with resetting used for modelling income dynamics and developing a framework that is able to crucially \textit{disaggregate} the temporal properties of income to the \textit{level of an individual worker}. We find analytically and illustrate numerically that our framework is orthogonal to the transition matrix approach and leads to improved and more granular estimates. Moreover, to facilitate empirical applications of the framework, we introduce a publicly available statistical methodology, and showcase the application using the USA income dynamics data. These results help to improve our understanding on the temporal properties of income in real economies and provide a set of tools for designing policy interventions.
\end{abstract}

\newpage


\section*{Introduction}\label{sec:intro}

\begin{figure}[H]
\centering
\includegraphics[width=8.7cm]{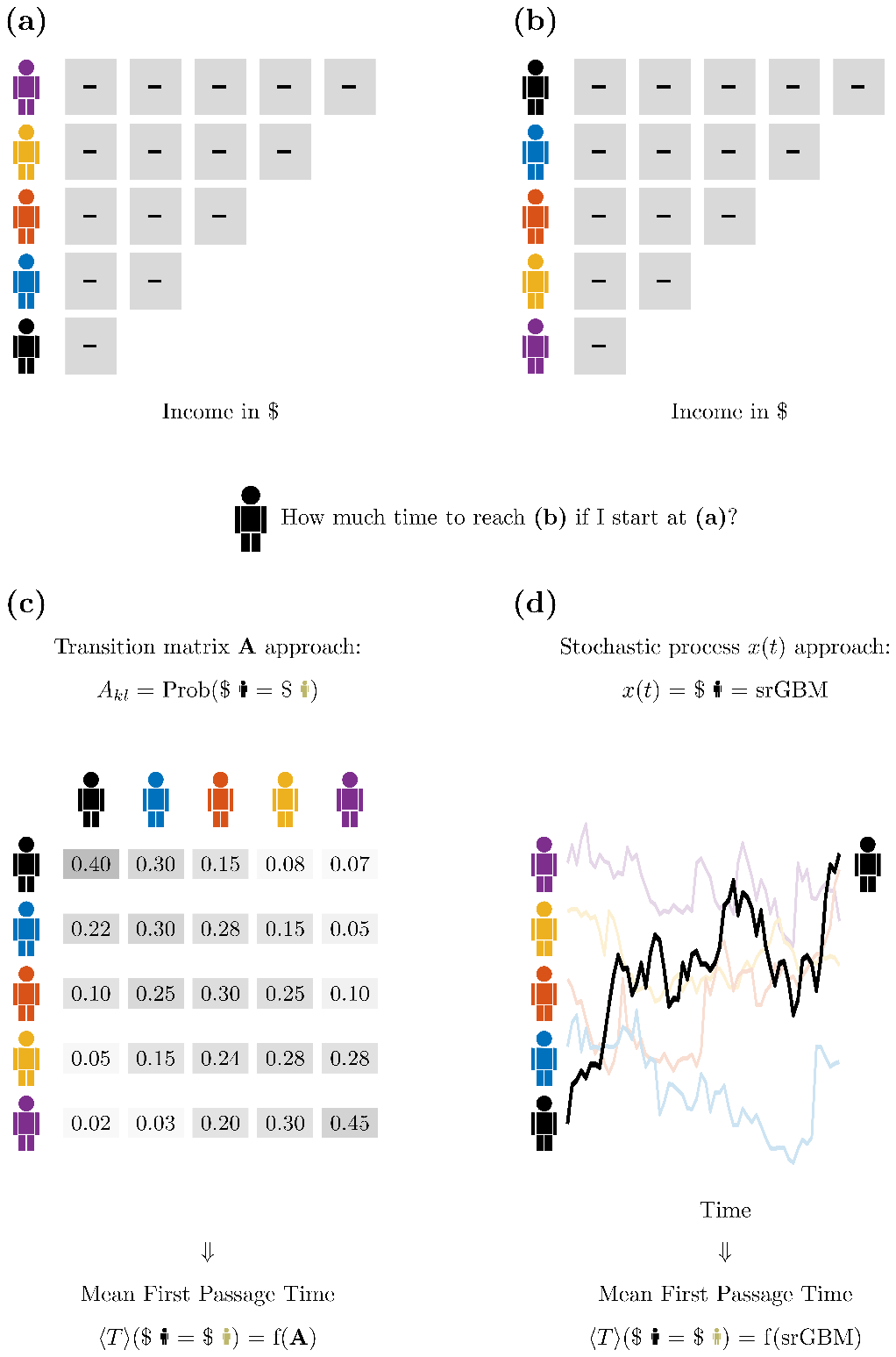}
\caption{\textbf{Approaches to measuring the time properties of income}: \textbf{(a)} Bar chart for the income within a population in an initial time period. \textbf{(b)} Bar chart for the income within a population in a target time period. In both \textbf{(a-b)} the workers are sorted according to their income in descending order. \textbf{(c)} Transition matrix $\mathbf{A}$ measuring the transition probabilities between income quantiles for the workers in \textbf{(b)} if they started in \textbf{(a)}. The entries of the matrix describe the probabilities to move between two quantiles, and the MFPT $\langle T\rangle$ is a function of $\mathbf{A}$. \textbf{(d)} In our approach, the income dynamics of a worker is a stochastic process the  MFPT $\langle T\rangle$ to reach the target income in \textbf{(b)} if a worker started in \textbf{(a)} is a function of parameters that quantify the state of the economy.}
\label{fig:contribution}
\end{figure}

How long does it take for individuals to improve their income profiles within the socio-economic ladder? For James Truslow Adams, this question formed the bedrock for the existence of the American dream~\cite{adams1931epic}. If the time is comparable to the working life of an individual, then it can be argued that social and economic systems function as envisaged. Otherwise, there might be a need to reorganize socio-economic policies. As a result, developing methods for studying the temporal properties of income dynamics is a fundamental question in the literature of economic inequality and mobility~\cite{fields1996meaning,corak2013income,solon1992intergenerational,shorrocks1982inequality,shorrocks1980class,shorrocks1983ranking,jantti2015income,shorrocks1978measurement} (Fig.~\ref{fig:contribution}(a-b)).

These temporal changes  are usually studied by using \textit{income transition matrices}~\cite{shorrocks1978measurement,jantti2015income,kreiner2018role,cheng2019americans,bulczak2022downward}. An income transition matrix aggregates and summarizes the probability of a worker to move between two arbitrary income quantiles, $k$ and $l$, within a given time period (Fig.~\ref{fig:contribution}(c)). Using this matrix, we can estimate the expected time for a worker to first reach the lowest income that is required to reach $l$ given that the worker is currently in quantile $k$. This quantity is known as the mean first passage time (MFPT)~\cite{redner2001guide,metzler2014first,pal2017first} and is an adequate approximation for the time required for a worker to improve their income level~\cite{metzler2014first,redner2001guide}.

Transition matrices, however, only provide aggregated quantities for the time properties of income. That is, because of the aggregation we are unable to differentiate the fortunes of workers that are members of the same quantile. Indeed, empirical observations of income dynamics exhibit a fractal like structure~\cite{pareto1964cours}, implying that not everyone within the quantile will have the same probability to change their income status. Instead, there are intra-quantile dependencies that govern the MFPT. As a result, in order to move away from this drawback of transition matrices, the more recent literature has developed alternate measures for tracking income fluctuations that are able to disaggregate the transition matrices to a higher resolution~\cite{bhattacharya2011nonparametric}. But, despite these innovations, our knowledge about the time at which these fluctuations occur remains limited.

To bridge this gap, we build an analytical framework for disaggregating the MFPT to the level of an individual worker in the economy. We exploit the properties of an established stochastic process used for modelling income dynamics called Geometric Brownian Motion with stochastic resetting (srGBM)~\cite{stojkoski2021geometric,stojkoski2022income}. srGBM has been widely used for investigating the role of various phenomena on income dynamics: from taxes~\cite{aoki2017zipf} up to changes in skill prices~\cite{gabaix2016dynamics}. Mathematically, our srGBM-MFPT framework can be seen as a natural generalization of the finite-state transition matrix approach to processes with a continuum of states (Fig.~\ref{fig:contribution}(d))~\cite{gabaix2016dynamics}. Practically, it can be seen as an answer to whether the American Dream is a reality for a particular worker. We utilize our analytical results to develop a statistical methodology for applying the MFPT in real-world data and display its application by providing estimates for the time properties of the income distribution in the United States (USA) for the period between 1978-2015. With these estimates we are able to see a more granular and comprehensive picture than previous approaches for the ability of a worker to move across the income distribution. Hence, they can be used by policy makers to answer important questions, such as: What is the time needed for a currently minimum wage worker to spend in the workforce in order to reach a reasonable income level? Which proportion of workers are able to reach the highest status during their working life? How easy it is for workers to reach certain income targets? How can we recommend policies for optimizing the time required for a worker to improve their status in an economy? To support research and policy interventions in the pursuit of the answers to these questions, we release a simplified notebook for calculating the srGBM MFPT by using real world data at: \url{https://colab.research.google.com/drive/1quV4bdaNaGVmUB1EA0XOINqWC93Y8JBX}.

\section*{Results}

To develop a ``disaggregated'' view for the MFPT across each worker of the income distribution, we use the properties of geometric Brownian motion with stochastic resetting (srGBM)~\cite{stojkoski2021geometric,vinod2022nonergodicity,stojkoski2022income,gripenberg1983stationary,gabaix2016dynamics,pakes1987limit,van2021unified,vinod2022time2} which is pertinent to income dynamics. For example, this model confers several real world socio-economic phenomena such as the power law stationary income distributions \cite{gabaix2016dynamics,nirei2004income,aoki2017zipf} or the famous Great Gatsby curve that visualizes the relationship between inequality and mobility~\cite{stojkoski2022income}. In our approach, we simply assume that time is continuous and there is a population of $N$ workers in the economy. The income $x(t)$ of a worker in period $t$ grows multiplicatively with a rate $\mu$ and volatility $\sigma$ until a random event that occurs with a rate $r$ resets its dynamics~\cite{evans2011diffusion}. The reset event can be interpreted as a worker that left the job market (for example by retiring, being laid off or after an injury) and is substituted by another younger worker with a starting income $x_r$~\cite{nirei2004income} Further details about srGBM can be found in the Supplementary Information (SI) Section~\ref{app:srGBM}.

\subsection*{MFPT in srGBM}

How long does it take for a worker having a certain initial income $x_0$ to reach a target threshold $y$ with intermittent resets to a new and fluctuating income $x_r$?
This time is random owing to the fact that the growth of income is inherently stochastic and furthermore there is a temporal stochasticity induced by resetting. The MFPT estimates the expected time for an ensemble of such processes~\cite{redner2001guide,metzler2014first,pal2017first} of reaching income $y$ when starting with $x_0$ and is given by (Section~\ref{app:srbm-mfpt-derivation})
\begin{equation}
\langle T_r(x_0,y,x_r) \rangle = \frac{1-\tilde{T}(x_0,y,r)}{r\, \tilde{T}(x_r,y,r)},
\label{eq:srgbm-mfpt}
\end{equation}
where 
\begin{align}
\tilde{T}(x_0,y,r)=\left( \frac{x_0}{y} \right)^{q_1}, \quad
 q_1&=\frac{\sqrt{(\sigma^2-2\mu)^2+8r \sigma^2}+(\sigma^2-2\mu)}{2 \sigma^2}.
\end{align}

We observe that larger growth of income (Fig.~\ref{fig:mfpt-theoretical}(a)) and larger randomness in the economy (Fig.~\ref{fig:mfpt-theoretical}(b)) decrease the MFPT in srGBM (see also SI Section~\ref{app:mfpt-srGBM} for numerical methods). These two notable factors indeed comply with typical observations, i.e., it is easier for workers to move across the income distribution when the economy is growing at faster rates and when there is more volatility~\cite{aristei2015drivers}. Furthermore, we observe that there is an optimal resetting rate ($r^*$) which minimizes the MFPT $\langle T_r \rangle$ in \eref{eq:srgbm-mfpt}. In economics terms, the resetting rate can be seen as a controlled factor that policy makers can control in order to optimize the dynamics of income within an economy. For instance, they can develop policies that are aimed at increasing/decreasing the number of workers who retire or leave their job as a means to reduce time required for a worker to move across the income distribution~\cite{cremer2004social,staubli2013does,sheather2021great}. In this context, we find two additional important facts that can help policy makers:  First, for a fixed volatility, larger growth ($\mu$) decreases the optimal rate at which the MFPT is at a minimum level (inset of Fig.~\ref{fig:mfpt-theoretical}(a)). Policies that lead to efficient qualification programs, or to quality foreign investments can, for example, be attributed to this effect. Second, for a fixed growth rate, when the randomness in the system is increased, we also observe an increased optimal resetting rate (inset of Fig.~\ref{fig:mfpt-theoretical}(b)). This, for example can be a result of increased intrinsic differences between societies when it comes to getting new jobs.

\subsection*{Transition matrix vs. Stochastic process approach for MFPT} 

Differently from our approach, the state of the art methods for studying income dynamics rely on utilizing the properties of the income transition matrix~\cite{JanttiJenkins2015}. But, as we will show here, transition matrices are in many ways inferior to our framework.

An income transition matrix aggregates the income movements over a given period in time~\cite{shorrocks1978measurement}. It summarizes the mobility in a
stochastic matrix $\mathbf{A}$ in which the elements $A_{kl}$ quantify the probability that an individual in income quantile $k$ in time $t$ is found in income quantile $l$ in time $t + \Delta$. Formally, let $\mathcal{S}_k(t)$ denote the set of individuals that are part of quantile $k$ in time $t$. Then,
\begin{align}
    A_{kl} = \frac{|\mathcal{S}_k(t) \cap \mathcal{S}_l(t+\Delta) | }{|\mathcal{S}_l(t+\Delta) |},
\end{align}
where $|\mathcal{S}|$ is the cardinality of $\mathcal{S}$~\cite{jantti2015income}. 

\begin{figure}[H]
\centering
\includegraphics[width=17cm]{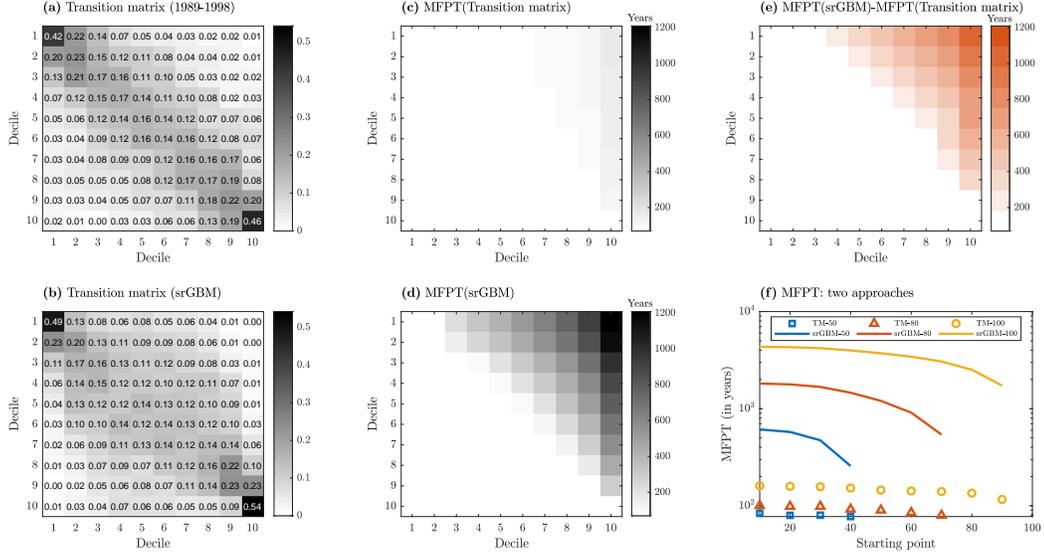}
\caption{\textbf{Transition matrix MFPT vs. srGBM MFPT}. \textbf{(a)}
 Income transition matrix in USA for the period 1989-1998, taken from~\cite{JanttiJenkins2015}. \textbf{(b)} Income transition matrix estimated through srGBM with parameters $\mu=0.10$ year$^{-1}$, $\sigma^2=0.03$ year$^{-1}$, $r=0.041$ year$^{-1}$, chosen to match the empirical transition matrix in \textbf{(a)}. This was done by minimizing the Frobenius norm of the difference matrix between the empirical and the one derived from srGBM. The srGBM transition matrix is able to explain 83\% of the variations in the original transition matrix. \textbf{(c)} MFPT (in years) between deciles using the transition matrix approach from the data in \textbf{(a)}. \textbf{(d)} MFPT (in years) between deciles calculated using the srGBM MFPT approach. \textbf{(e)} Difference between the MFPT obtained from the transition matrix \textbf{(c)} and srGBM approaches \textbf{(d)}. \textbf{(f)} Also shows the difference between the two approaches by investigating the MFPT to a target income as a function of the starting income.}
\label{fig:mfpt-srgbm-matrix}
\end{figure}

In a perfectly mobile economy, the entries of the transition matrix are all equal to each other. Realistic income transition matrices, however, are characterized with higher mobility at the bottom of quantiles than at the top. To illustrate this, in Fig.~\ref{fig:mfpt-srgbm-matrix}(a)  we display the income transition matrix for the United States (for the period between 1989-1998) by using data taken from Ref.~\cite{JanttiJenkins2015}). In Fig.~\ref{fig:mfpt-srgbm-matrix}(b) we show that the income transition matrix in srGBM can easily reproduce these properties by conducting a simple experiment. The experiment is based on generating an artificial economy with a population of one million workers. Each worker is assigned an initial income that is drawn from the stationary srGBM distribution with parameters that best fit the real income transition matrix (Fig.~\ref{fig:mfpt-srgbm-matrix}(a)). The dynamics of the income is then simulated for a sufficiently long time and the mobility dynamics across a 10 year period are aggregated into deciles. The srGBM generated transition matrix is able to explain 83\% of the variations in the real transition matrix, thus suggesting that the model indeed adequately reproduces the real world income mobility.

Using the information contained in the transition matrix, we can quantify the MFPT for an individual to move between two quantiles $k$ and $l$ using various methods (see SI Section~\ref{app:mfpt-matrices} and Refs.~\cite{israel2001finding,hunter2018computation,fronczak2009biased,noh2004random}).

Before we detail the similarities and differences between the srGBM and the transition matrix MFPT (TMFPT), several of its features need to be pointed out. First, the TMFPT is measured in $\Delta$ periods (e.g., in the cases described in Fig.~\ref{fig:mfpt-srgbm-matrix}(a-b), the MFPT is measured in decades), but unless there are changes in the economic conditions, its value normalized in a different unit (e.g., years) is independent of $\Delta$.
Second, the income transition matrix is ignorant of the growth, volatility, and resetting rates. Instead, these parameters are bulked together in the transition probabilities. Lastly, the majority of the methods for estimation of the TMFPT rely upon discretized time dynamics, and thus need to be modified to capture the continuous time dynamics~\cite{israel2001finding}. Hence, every estimation based on the transition matrix can only be \textit{an approximation} for the MFPT.

The relationship between the srGBM-MFPT and the TMFPT can be shown in two different ways. First, for TMFPT between any two quantiles, we can estimate the starting income and the target income that generate the same srGBM-MFPT. Intuitively, the starting income belongs to the lower quantile and the target income to the upper quantile, but in general the position of the starting income depends on the extent of aggregation. Second, we can expand the number ($K$) of quantiles of the transition matrix until they reach the total number ($N$) of workers in the economy (and suppose that $N$ is sufficiently large). This is also known as the finite state to continuum approach in statistical physics~\cite{gabaix2016dynamics}. The limiting case as $N\to\infty$ and $K\to N$ results in the srGBM-MFPT. 

In general, the aggregation of real world income dynamics leads to underestimating the MFPT (see Figs.~\ref{fig:mfpt-srgbm-matrix}(c-f), for various comparisons between srGBM-MFPT and TMFPT). This is due to the fact that the aggregation leads to neglecting the differences in the income distribution that appear within a quantile. That is, income within a quantile is known to have a fractal like structure, i.e., the income distribution within a quantile also follows a power law~\cite{pareto1964cours}. This implies that not everyone within the quantile will have the same probability to change their income status. Instead, there are intra-quantile dependencies that govern the MFPT. The TMFPT is unable to identify this phenomenon because the aggregation essentially removes certain parts of the income distribution and quantifies income dynamics based on quantile positions. In stark contrast, the srGBM-MFPT captures this distribution dependence because it models the exact starting and the target income that a worker wants to reach.

\subsection*{Applying the srGBM MFPT to real world data}

We can apply the presented theoretical framework to answer real world questions that form the basis for developing socio-economic policies, such as: What is the time needed for a currently minimum wage worker to spend in the workforce in order to reach a reasonable income level? Which workers are able to reach the highest status during their working life? How easy it is for workers to reach certain income targets? How can we optimize the time required for a worker to improve their status in an economy? 


To answer these questions, we develop a statistical methodology for estimating the MFPT across two points of the income distribution. Our methodology, in general, relies on two assumptions which can be relaxed depending on the availability of data. First, we assume that the srGBM parameters can be discretized and are able to change each year. This reflects changes in economic conditions and policies. Then, the srGBM parameters can be estimated directly using data for the dynamics of the income distribution or taken from other sources. Because these are only estimates for the srGBM parameters, they are coupled with confidence intervals for the MFPT (See SI Section~\ref{app:srgbm-empirical-methodology} and Refs.~\cite{gabaix2016dynamics,aoki2017zipf,stojkoski2022income}). Second, we assume that the resetting rate can be approximated using data on the working age population who lost and/or left their job within a calendar year. In what follows, we will display results for the MFPT between percentiles of the income distribution, but we point out that the results can be easily disaggregated into an even higher resolution.


Let us now show how we can use the data from~\cite{stojkoski2022income} and our srGBM MFPT method to estimate confidence intervals for the required time for US workers to move between two arbitrary percentiles of the income distribution (See SI Section~\ref{app:empirical-srgbm-MFPT} for more details about the statistical estimation). We will thereby assume that there are no changes in the economic conditions once we estimate the MFPT. That is, we will answer questions about the time properties of income for each of the studied years under the assumption that the economy will remain the same over the estimated MFPT. Thus, any changes in the MFPT over the years can be attributed to changes in the economic policies or external shocks to the economy.

\paragraph{What is the required time for a low income worker to reach a reasonable income?}\label{time-leave}

\begin{figure}[H]
\centering
\includegraphics[width=14cm]{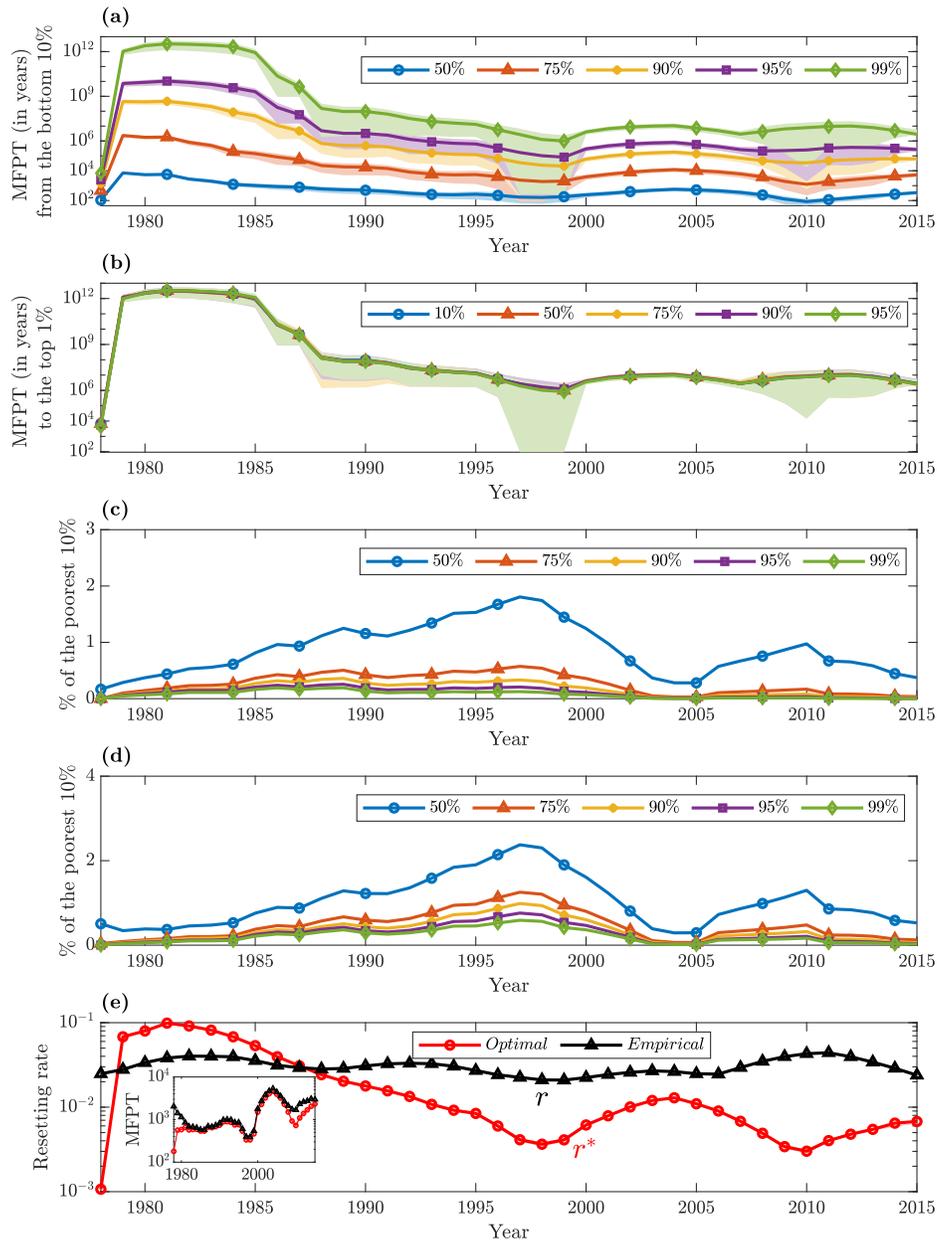}
\caption{\textbf{MFPT in the US income distribution as a function of time}. In the top panel we show \textbf{(a)} The MFPT for a worker in the bottom 10\% to reach 50\%, 75\%, 90\%, 95\% and 99\%.  \textbf{(b)} The MFPT for reaching 99\% of the income distribution for a worker belonging in 10\%, 50\%, 75\%, 90\% and 95\%. In the bottom panel we show the fraction of poorest 10\% which reach the income of the 50\%, 75\%, 90\%, 95\% and 99\% in \textbf{(c)} 20 years, \textbf{(d)} 40 years, \textbf{(e)} Empirical resetting rate and mean optimal resetting rate as a function of time. The inset plot gives the corresponding MFPT for the optimal and the empirical resetting rate. The color filled regions represent two standard error confidence interval bands. The srGBM parameters are the same as those estimated in Ref.~\cite{stojkoski2022income}. }\label{fig:US-mfpt}
\end{figure}

In Fig.~\ref{fig:US-mfpt}(a) we show the estimated MFPT for a worker to leave the lowest 10\% and reach the level of income of the 50\%, 75\%, 90\%, 95\% and 99\%, given that they started at the middle position in the decile. We observe that the MFPT of all these targets was decreasing and until the end of the 1990s, when they reached their minimum.  At the end of the 1990s these MFPTs drastically increased and stayed at a high level until the last year of our analysis (2015). 

\paragraph{Which workers are able to reach the highest income status during their working life? 
}\label{time-top1}

Next, we investigate how long it would take a person to reach the income levels of the 99 percentile, given that they start at an average position in various quantiles of the income distribution (Fig.~\ref{fig:US-mfpt}(b)).

Similar to the previous results, the amount of time needed to reach the highest income status was at its lowest values in the end of the 1990s. However, the MFPT values were much larger compared to the working life of an individual. For instance, in these periods, the workers which started in the 95th percentile of the income distribution, had to work around 700 years in 1999 to reach the highest paid percentile, though with a very wide confidence interval. This MFPT, remained at a consistently high level until the last year of our analysis.

\paragraph{How easy it is for workers to reach certain income targets? }\label{fractions}

We can also analyze the fraction of the lowest income workers that constitute the bottom decile which reach a certain target in the income distribution in a time span that corresponds to the duration of a typical working life. For this we can reverse the MFPT equation and estimate how many workers are able to reach an MFPT within a given time span of X years (see SI Section~\ref{app:srgbm-empirical-fraction} for the mathematical details).

For example, the economic conditions in the US allowed only less than 2\% of workers coming from the lowest income level to reach the highest income status in a span of 20 years ever since 1978. This percentage drops significantly as we move to more recent years and to larger income targets (Fig.~\ref{fig:US-mfpt}(c)). The percentage becomes not much larger, if we set X to a larger value, such as 40 years (Fig.~\ref{fig:US-mfpt}(d)).

\paragraph{How can we optimize the MFPT?}\label{real-optimal}

Finally, we can ask ourselves how close is the estimated MFPT to the optimal MFPT and what can policy makers do to optimize it?

In Fig.~\ref{fig:US-mfpt}(e) we plot the mean optimal resetting rate and compare it with the empirical resetting rate (See SI Section~\ref{app:srgbm-empirical-optimal-rate} on the methods of estimation). The inset plot compares the optimal MFPT and the one observed in reality. We find that the empirical resetting rate is persistently above the optimal, starting from 1988. In other words, it should be easily possible to decrease the MFPTs of the US income distribution by developing policies that decrease the resetting rate.

\section*{Conclusion and Discussion}

Socio-economic policies aimed at improving the welfare of individuals often rely upon estimates for the typical time-frame required for workers to improve their income status. Yet, state-of-the-art methods for these estimates can only offer an aggregated view. Here, we developed a disaggregate measure by exploiting the MFPT, a simple estimate for this time, in the srGBM baseline model for income dynamics. We found that the srGBM-MFPT adequately reproduces the real world features of income dynamics, often depicted in an income transition matrix. We showed that the srGBM-MFPT is an orthogonal measure to standard metrics. It provides a deeper insight into the time properties of income within an economy as it is able to uncover the workers which are able to move across the whole space of the income distribution during their working life. This includes moves within an aggregated single level of the transition matrix approach. Also, we created an empirical methodology for applying our theoretical results. We showed that the methodology is equipped with a set of tools which make it capable of answering specific questions about the time properties of income for any economic structure, and presented its application using US income data. Hence, we expect it to be of wide interest to practitioners for developing policy interventions aimed at optimizing the time. To facilitate further research and policy interventions, we share a simplified code for calculating the srGBM-MFPT at \url{https://colab.research.google.com/drive/1quV4bdaNaGVmUB1EA0XOINqWC93Y8JBX}.

We point out certain limitations to our approach. First, these estimates are based on srGBM which can offer a good but nevertheless information about the income dynamics. Our framework assumes that all workers have the same socio-economic attributes that may affect the evolution of income, such as education or gender, and does not differentiate across different types~\cite{aydemir2013intergenerational,corak2006poor,haveman2006role,jacobs1996gender,jacobs1996gender,reay2018miseducation,troyna2012racial,klasen2009impact}. Instead, it captures the attributes of workers by two parameters: the growth of income and the volatility. We emphasize that the methodology can be  generalized and applied to more sophisticated situations by assuming that the growth rates and volatilities are type dependent or when they are themselves random variables (e.g., via doubly-stochastic models)~\cite{gabaix2016dynamics}. Then, each worker type will have their own MFPT (estimated using the same equation). This may uncover even more detailed information about the time properties of the income distribution and considerably enhance the  interpretation of our results. Indeed, this appears to be fruitful topic for expanding the srGBM-MFPT framework. 

Second, here we studied only one aspect of socio-economic inequality, the income distribution and its evolution. The American dream is also related to wealth inequality and its dynamics, two aspects that are yet not captured by our framework. Indeed, recent studies have suggested that the time properties of the realistic wealth distribution may not exhibit the features which are usually assumed in policy practices~\cite{berman2020wealth,stojkoski2022ergodicity}. Hence, extending the framework to account for wealth may be a non-trivial future contribution. 

Third, we restricted our empirical analysis to only one economy. We do not go into a detailed investigation on the geographical distribution of srGBM, which might be essential for understanding the spatial differences in various socio-economic phenomena.

Lastly, we emphasize that the application which we presented here is only descriptive. We invite policy practitioners to delve deeper into the interpretation of the questions arising from our empirical findings. That is: Is the American Dream alive if it takes around 70 years for a middle-income status US worker to reach the top 1\%? Is it reasonable that the economic conditions allow only 1\% of the lowest income workers to reach the highest income rank? What were the reasons for the sub-optimal resetting rate over the years?  In this context, the MFPT can provide much more information than the extent of income mobility within an economy: it is able to uncover the feasibility of each worker to move across the income distribution~\cite{stojkoski2022measures}.

Yet, despite these limitations, the srGBM MFPT framework significantly improves upon the state of the art by providing a method that is more comprehensive, and also more accurate, at investigating the dynamics of a worker's income. This methodology advances our understanding of the time dimension of an individual worker's income, and improves the current state-of-the-art. It will motivate new multidisciplinary research focused on creating even more comprehensive methods that can be used to predict the time properties of income -- mean and higher order metrics -- in specific domains of the economy and be applied to economies all across the globe.

\section*{Materials and methods}

\subsection*{Geometric Brownian Motion with stochastic resetting}

We denote the income of a worker at time $t$ by $x(t)$. As explained in the main text, the income dynamics empirically follows srGBM and can be described by the following stochastic equation
\begin{align}
d x(t) &=(1-Z_{t}) x(t) \left[ \mu dt+ \sigma dW \right]+Z_{t} \left( x_0-x(t) \right),
\label{eq:srgbm-microscopic}
\end{align}
where $dt$ denotes the infinitesimal time increment and
$dW$ is an infinitesimal Wiener increment, which is normal-variate with $\langle dW_t \rangle=0$ and $\langle dW_t dW_s \rangle =\delta(t-s)dt$. Here $\delta(t)$ denotes the Dirac $\delta$-function. Resetting is implemented with a random variable $Z_t$ which resets the income dynamics to the initial value $x(0) = x_r =$ minimum observed income. To be specific, 
$Z_t$ takes the value $1$ when there is a resetting event in the time interval between $t$ and $t+dt$; otherwise, it remains zero. For simplicity, we assume that the probability for a reset event, i.e., $\text{Prob}(Z_{t} = 1)$ in the interval $dt$ is given by $rdt$. In the limit $dt \to 0$, this confers to an exponential waiting time density, namely $re^{-rt}$ for resetting events. Notice that resetting can take the value of income $x(t)$ to go below $x_0$, but it cannot make it negative.

\subsection*{Empirical values of srGBM parameters in USA}

We use the yearly estimates for the parameters needed to estimate the srGBM provided in~\cite{stojkoski2022income} (see also SI Section~\ref{app:srgbm-empirical-methodology}). Moreover, each year, the true resetting rate is approximated with the fraction of the working age population (15-64) in the USA who lost and/or left their job within a calendar year. Workers who lost their job are those that either are temporarily laid off as well as those who permanently lost their jobs. The job leavers, on the other hand, are those that quit and immediately began searching for a new work. We take these data from the dataset for unemployment provided by the U.S. Bureau of Labor Statistics. The time series covers the period from 1977 up to 2015 and can be accessed at~\url{ https://fred.stlouisfed.org}. 

\subsection*{Estimating the srGBM MFPT in the USA}


In order to study the first passage times in the US income distribution we need to specify an initial position and an absorbing boundary. 
We estimate these quantities by providing a starting percentile and a target percentile, then we translate these values to the corresponding initial position, and target income in the estimated income distribution, which is calculated by running an srGBM simulation of a model economy, using the fitted empirical values of the model parameters. 

To be more specific, if we want to estimate the average time span before a typical worker of the bottom percentile reaches a top percentile, then we fix their threshold percentile income as a starting position of the stochastic process (srGBM) and the target percentile ``entry level'' income as a virtual absorbing boundary. The minimum observed income within the population is set as the resetting point.

\section*{Acknowledgments}
The authors acknowledge financial support by the German Science Foundation (DFG, Grant No. ME 1535/12-1). Arnab Pal gratefully acknowledges research support from the Department of Science and Technology, India, SERB Start-up Research Grant Number SRG/2022/000080 and Department of Atomic Energy, India. Trifce Sandev was supported by the~Alexander von~Humboldt Foundation.


\begin{thebibliography}{10}

\bibitem{adams1931epic}
James~Truslow Adams.
\newblock {\em The Epic of America}.
\newblock Routledge, 1931.

\bibitem{fields1996meaning}
Gary~S Fields and Efe~A Ok.
\newblock The meaning and measurement of income mobility.
\newblock {\em J. Econ. Theory}, 71(2):349--377, 1996.

\bibitem{corak2013income}
Miles Corak.
\newblock Income inequality, equality of opportunity, and intergenerational
  mobility.
\newblock {\em J. Econ. Perspect.}, 27(3):79--102, 2013.

\bibitem{solon1992intergenerational}
Gary Solon.
\newblock Intergenerational income mobility in the united states.
\newblock {\em Am. Econ. Rev.}, 82:393--408, 1992.

\bibitem{shorrocks1982inequality}
Anthony~F Shorrocks.
\newblock Inequality decomposition by factor components.
\newblock {\em Econometrica}, 50:193--211, 1982.

\bibitem{shorrocks1980class}
Anthony~F Shorrocks.
\newblock The class of additively decomposable inequality measures.
\newblock {\em Econometrica}, 48:613--625, 1980.

\bibitem{shorrocks1983ranking}
Anthony~F Shorrocks.
\newblock Ranking income distributions.
\newblock {\em Economica}, 50(197):3--17, 1983.

\bibitem{jantti2015income}
Markus J{\"a}ntti and Stephen~P Jenkins.
\newblock Income mobility.
\newblock In {\em Handbook of income distribution}, volume~2, pages 807--935.
  Elsevier, 2015.

\bibitem{shorrocks1978measurement}
Anthony~F Shorrocks.
\newblock The measurement of mobility.
\newblock {\em Econometrica}, 46:1013--1024, 1978.

\bibitem{kreiner2018role}
Claus~Thustrup Kreiner, Torben~Heien Nielsen, and Benjamin~Ly Serena.
\newblock Role of income mobility for the measurement of inequality in life
  expectancy.
\newblock {\em Proceedings of the National Academy of Sciences},
  115(46):11754--11759, 2018.

\bibitem{cheng2019americans}
Siwei Cheng and Fangqi Wen.
\newblock Americans overestimate the intergenerational persistence in income
  ranks.
\newblock {\em Proceedings of the National Academy of Sciences},
  116(28):13909--13914, 2019.

\bibitem{bulczak2022downward}
Grzegorz Bulczak and Alexi Gugushvili.
\newblock Downward mobility among individuals with poor initial health is
  linked with higher cardiometabolic risk in the united states.
\newblock {\em PNAS Nexus}, 2022.

\bibitem{redner2001guide}
Sidney Redner.
\newblock {\em A guide to first-passage processes}.
\newblock Cambridge University Press, 2001.

\bibitem{metzler2014first}
Ralf Metzler, Sidney Redner, and Gleb Oshanin.
\newblock {\em First-passage phenomena and their applications}, volume~35.
\newblock World Scientific, 2014.

\bibitem{pal2017first}
Arnab Pal and Shlomi Reuveni.
\newblock First passage under restart.
\newblock {\em Phys. Rev. Lett.}, 118(3):030603, 2017.

\bibitem{pareto1964cours}
Vilfredo Pareto.
\newblock {\em Cours d'{\'e}conomie politique}, volume~1.
\newblock Librairie Droz, 1964.

\bibitem{bhattacharya2011nonparametric}
Debopam Bhattacharya and Bhashkar Mazumder.
\newblock A nonparametric analysis of black--white differences in
  intergenerational income mobility in the united states.
\newblock {\em Quantitative Economics}, 2(3):335--379, 2011.

\bibitem{stojkoski2021geometric}
Viktor Stojkoski, Trifce Sandev, Ljupco Kocarev, and Arnab Pal.
\newblock Geometric brownian motion under stochastic resetting: A stationary
  yet nonergodic process.
\newblock {\em Phys. Rev. E}, 104(1):014121, 2021.

\bibitem{stojkoski2022income}
Viktor Stojkoski, Petar Jolakoski, Arnab Pal, Trifce Sandev, Ljupco Kocarev,
  and Ralf Metzler.
\newblock Income inequality and mobility in geometric brownian motion with
  stochastic resetting: theoretical results and empirical evidence of
  non-ergodicity.
\newblock {\em Philos. Trans. R. Soc. A}, 380(2224):20210157, 2022.

\bibitem{aoki2017zipf}
Shuhei Aoki and Makoto Nirei.
\newblock Zipf's law, pareto's law, and the evolution of top incomes in the
  united states.
\newblock {\em Am. Econ. J. Macroecon.}, 9(3):36--71, 2017.

\bibitem{gabaix2016dynamics}
Xavier Gabaix, Jean-Michel Lasry, Pierre-Louis Lions, and Benjamin Moll.
\newblock The dynamics of inequality.
\newblock {\em Econometrica}, 84(6):2071--2111, 2016.

\bibitem{vinod2022nonergodicity}
Deepak Vinod, Andrey~G Cherstvy, Wei Wang, Ralf Metzler, and Igor~M Sokolov.
\newblock Nonergodicity of reset geometric brownian motion.
\newblock {\em Phys. Rev. E}, 105(1):L012106, 2022.

\bibitem{gripenberg1983stationary}
G~Gripenberg.
\newblock A stationary distribution for the growth of a population subject to
  random catastrophes.
\newblock {\em Journal of Mathematical Biology}, 17(3):371--379, 1983.

\bibitem{pakes1987limit}
Anthony~G Pakes.
\newblock Limit theorems for the population size of a birth and death process
  allowing catastrophes.
\newblock {\em Journal of Mathematical Biology}, 25(3):307--325, 1987.

\bibitem{van2021unified}
Remco van~der Hofstad, Stella Kapodistria, Zbigniew Palmowski, and Seva Shneer.
\newblock Unified approach for solving exit problems for additive-increase and
  multiplicative-decrease processes.
\newblock {\em arXiv preprint arXiv:2102.00438}, 2021.

\bibitem{vinod2022time2}
Deepak Vinod, Andrey~G Cherstvy, Ralf Metzler, and Igor~M Sokolov.
\newblock Time-averaging and nonergodicity of reset geometric brownian motion
  with drift.
\newblock {\em Physical Review E}, 106(3):034137, 2022.

\bibitem{nirei2004income}
Makoto Nirei and Wataru Souma.
\newblock Income distribution and stochastic multiplicative process with reset
  event.
\newblock In {\em The Complex Dynamics of Economic Interaction}, pages
  161--168. Springer, 2004.

\bibitem{evans2011diffusion}
Martin~R Evans and Satya~N Majumdar.
\newblock Diffusion with stochastic resetting.
\newblock {\em Phys. Rev. Lett.}, 106(16):160601, 2011.

\bibitem{aristei2015drivers}
David Aristei and Cristiano Perugini.
\newblock The drivers of income mobility in europe.
\newblock {\em Economic Systems}, 39(2):197--224, 2015.

\bibitem{cremer2004social}
Helmuth Cremer, Jean-Marie Lozachmeur, and Pierre Pestieau.
\newblock Social security, retirement age and optimal income taxation.
\newblock {\em J. Public Econ.}, 88(11):2259--2281, 2004.

\bibitem{staubli2013does}
Stefan Staubli and Josef Zweim{\"u}ller.
\newblock Does raising the early retirement age increase employment of older
  workers?
\newblock {\em J. Public Econ.}, 108:17--32, 2013.

\bibitem{sheather2021great}
Julian Sheather and Dubhfeasa Slattery.
\newblock The great resignation—how do we support and retain staff already
  stretched to their limit?
\newblock {\em BMJ}, 375(2533), 2021.

\bibitem{JanttiJenkins2015}
Markus J{\"a}ntti and Stephen~P. Jenkins.
\newblock Income mobility.
\newblock In {\em Handbook of Income Distribution}, volume~2, pages 807--935.
  Elsevier, 2015.

\bibitem{israel2001finding}
Robert~B Israel, Jeffrey~S Rosenthal, and Jason~Z Wei.
\newblock Finding generators for markov chains via empirical transition
  matrices, with applications to credit ratings.
\newblock {\em Mathematical finance}, 11(2):245--265, 2001.

\bibitem{hunter2018computation}
Jeffrey~J Hunter.
\newblock The computation of the mean first passage times for markov chains.
\newblock {\em Linear Algebra Appl.}, 549:100--122, 2018.

\bibitem{fronczak2009biased}
Agata Fronczak and Piotr Fronczak.
\newblock Biased random walks in complex networks: The role of local navigation
  rules.
\newblock {\em Phys. Rev. E}, 80(1):016107, 2009.

\bibitem{noh2004random}
Jae~Dong Noh and Heiko Rieger.
\newblock Random walks on complex networks.
\newblock {\em Phys. Rev. Lett.}, 92(11):118701, 2004.

\bibitem{aydemir2013intergenerational}
Abdurrahman Aydemir, Wen-Hao Chen, and Miles Corak.
\newblock Intergenerational education mobility among the children of canadian
  immigrants.
\newblock {\em Can. Public Policy}, 39(Supplement 1):S107--S122, 2013.

\bibitem{corak2006poor}
Miles Corak.
\newblock Do poor children become poor adults? lessons from a cross-country
  comparison of generational earnings mobility.
\newblock In {\em Dynamics of inequality and poverty}. Emerald Group Publishing
  Limited, 2006.

\bibitem{haveman2006role}
Robert Haveman and Timothy Smeeding.
\newblock The role of higher education in social mobility.
\newblock {\em Future Child.}, 16:125--150, 2006.

\bibitem{jacobs1996gender}
Jerry~A Jacobs.
\newblock Gender inequality and higher education.
\newblock {\em Ann. Rev. Sociol.}, 22:153--185, 1996.

\bibitem{reay2018miseducation}
Diane Reay.
\newblock Miseducation: Inequality, education and the working classes.
\newblock {\em Int. Stud. Sociol. Educ.}, 27(4):453--456, 2018.

\bibitem{troyna2012racial}
Barry Troyna.
\newblock {\em Racial inequality in education}.
\newblock Taylor \& Francis, 2012.

\bibitem{klasen2009impact}
Stephan Klasen and Francesca Lamanna.
\newblock The impact of gender inequality in education and employment on
  economic growth: new evidence for a panel of countries.
\newblock {\em Fem. Econ.}, 15(3):91--132, 2009.

\bibitem{berman2020wealth}
Yonatan Berman, Ole Peters, and Alexander Adamou.
\newblock Wealth inequality and the ergodic hypothesis: Evidence from the
  united states.
\newblock {\em Forthcoming in J. Income Distrib.}, 2020.

\bibitem{stojkoski2022ergodicity}
Viktor Stojkoski and Marko Karbevski.
\newblock Ergodicity breaking in wealth dynamics: The case of reallocating
  geometric brownian motion.
\newblock {\em Phys. Rev. E}, 105(2):024107, 2022.

\bibitem{stojkoski2022measures}
Viktor Stojkoski.
\newblock Measures of physical mixing evaluate the economic mobility of the
  typical individual.
\newblock {\em arXiv preprint arXiv:2205.02800}, 2022.

\bibitem{evans2020stochastic}
Martin~R Evans, Satya~N Majumdar, and Gr{\'e}gory Schehr.
\newblock Stochastic resetting and applications.
\newblock {\em J. Phys. A: Math. Theor.}, 53(19):193001, 2020.

\bibitem{pal2015diffusion}
Arnab Pal.
\newblock Diffusion in a potential landscape with stochastic resetting.
\newblock {\em Phys. Rev. E}, 91(1):012113, 2015.

\bibitem{dahlenburg2021stochastic}
Marcus Dahlenburg, Aleksei~V Chechkin, Rina Schumer, and Ralf Metzler.
\newblock Stochastic resetting by a random amplitude.
\newblock {\em Phys. Rev. E}, 103(5):052123, 2021.

\bibitem{vinod2022time}
Deepak Vinod, Andrey~G Cherstvy, Ralf Metzler, and Igor~M Sokolov.
\newblock Time-averaging and nonergodicity of reset geometric brownian motion
  with drift.
\newblock {\em Physical Review E}, 106(3):034137, 2022.

\bibitem{aitchison1957lognormal}
John Aitchison and James~AC Brown.
\newblock {\em The lognormal distribution with special reference to its uses in
  economics}.
\newblock Cambridge University Press, 1957.

\bibitem{stojkoski2020generalised}
Viktor Stojkoski, Trifce Sandev, Lasko Basnarkov, Ljupco Kocarev, and Ralf
  Metzler.
\newblock Generalised geometric brownian motion: Theory and applications to
  option pricing.
\newblock {\em Entropy}, 22(12):1432, 2020.

\bibitem{schiff1999laplace}
Joel~L Schiff.
\newblock {\em The Laplace transform: theory and applications}.
\newblock Springer Science \& Business Media, 1999.

\bibitem{eliazar2020power}
Iddo Eliazar.
\newblock {\em Power Laws}.
\newblock Springer, 2020.

\bibitem{aydiner2019money}
Ekrem Aydiner, Andrey~G Cherstvy, and Ralf Metzler.
\newblock Money distribution in agent-based models with position-exchange
  dynamics: the pareto paradigm revisited.
\newblock {\em Eur. Phys. J. B}, 92(5):1--4, 2019.

\bibitem{pal2016diffusion}
Arnab Pal, Anupam Kundu, and Martin~R Evans.
\newblock Diffusion under time-dependent resetting.
\newblock {\em J. Phys. A: Math. Theor.}, 49(22):225001, 2016.

\bibitem{reuveni2016optimal}
Shlomi Reuveni.
\newblock Optimal stochastic restart renders fluctuations in first passage
  times universal.
\newblock {\em Phys. Rev. Lett.}, 116(17):170601, 2016.

\bibitem{bonomo2022mitigating}
Ofek~Lauber Bonomo, Arnab Pal, and Shlomi Reuveni.
\newblock Mitigating long queues and waiting times with service resetting.
\newblock {\em PNAS Nexus}, 1(3):pgac070, 2022.

\bibitem{pal2019landau}
Arnab Pal and VV~Prasad.
\newblock Landau-like expansion for phase transitions in stochastic resetting.
\newblock {\em Phys. Rev. Research}, 1(3):032001, 2019.

\end{thebibliography}

\newpage

\newpage
\begin{figure}[H]
\centering
\includegraphics[width=15cm]{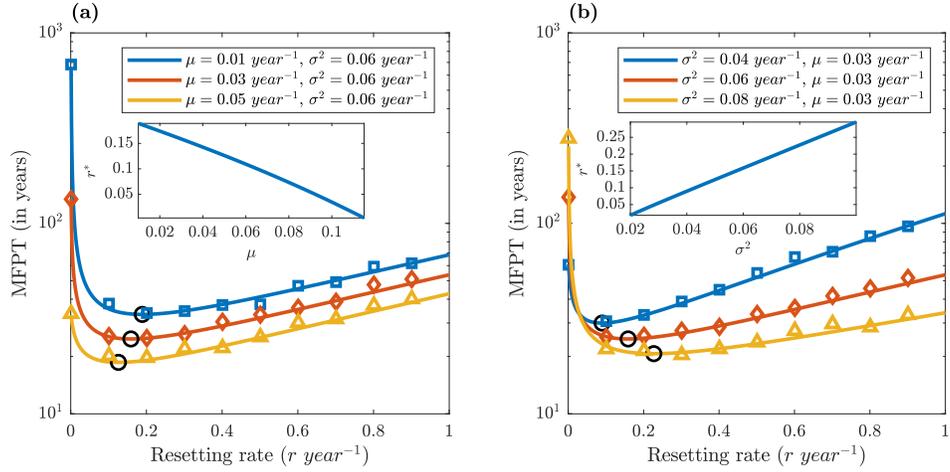}
\caption{\textbf{Dependence of the MFPT in srGBM on the model parameters} \textbf{(a)} MFPT in srGBM as a function of the resetting rate for various drift rates $\mu$. (b) Same as \textbf{(a)}, only for various volatilities $\sigma^2$. The black hollow circles show the optimal resetting rate. The inset plots show how the optimal resetting rate evolves as a function of $\mu$ and $\sigma^2$. More details about this optimal behavior can be found in SI Section~\ref{app:srbm-mfpt-derivation}. 
}\label{fig:mfpt-theoretical}
\end{figure}

\newpage

\newpage

\newpage

\newpage

\section*{Supplementary Information}

\setcounter{equation}{0}
\setcounter{figure}{0}
\setcounter{table}{0}
\setcounter{theorem}{0}
\setcounter{subsection}{0}
\setcounter{subsubsection}{0}
\makeatletter
\renewcommand{\theequation}{S\arabic{equation}}
\renewcommand{\thesubsection}{S\arabic{subsection}}
\renewcommand{\thetable}{S\arabic{table}}
\renewcommand{\thefigure}{S\arabic{figure}}
\renewcommand{\thetheorem}{S\arabic{theorem}}
\renewcommand{\theproposition}{S\arabic{proposition}}

\subsection{Properties of srGBM}\label{app:srGBM}

In the literature, resetting is described as an approximation for the external forces that influence the income dynamics and ensures a stationary distribution~\cite{evans2011diffusion,evans2020stochastic,pal2015diffusion,dahlenburg2021stochastic,vinod2022time}. We can use the renewal approach (see \cite{evans2020stochastic}) to show that that the probability density function (PDF) corresponding to $x(t)$ has a stationary solution. 
In particular, the PDF with resetting $(r>0)$ can be written as 
\begin{align}\label{eq:pdf-solution resetting}
  P_r(x,t|x_0,x_r) &= e^{-rt}P_{0}(x,t|x_0)+r\int_{0}^{t}e^{-ru}P_{0}(x,u|x_r)\,du,
\end{align}
where  $P_{r/0}(x,t|x_0/x_r)$ is the PDF of the reset/reset-free income dynamics. The reset-free PDF is a log-normal function (following It\^{o} convention) and reads~\cite{aitchison1957lognormal,stojkoski2020generalised}
\begin{align}
P_{0}(x,t|x_0)&= \frac{1}{x\sqrt{2\pi \sigma^2 t}}  \exp \left( \frac{-\left[\log (\frac{x}{x_0})-(\mu-\frac{\sigma^2}{2})t\right]^2}{2\sigma^2 t}  \right).
\end{align}
Finding the PDF $P_{r}(x,t|x_0,x_r)$ at all times usually is a daunting task, however a large time limit which is independent on the initial condition, $x_0$, can be obtained by making use of the final value theorem~\cite{schiff1999laplace}. Namely,
\begin{align}
    P_r^{ss}(x|x_r)=\lim_{t \to \infty} P_r(x,t|x_r)=\lim_{s \to 0} s\hat{P}_r(x,t|x_r)=r \hat{P}_{0}(x,r|x_r),
\end{align}
where $\hat{f}(s)=\mathcal{L}[f(t)]=\int_{0}^{\infty}e^{-st}f(t)\,dt$ is the Laplace transform of the function $f(t)$. Substituting Eq.~(\ref{eq:pdf-solution resetting}) into the above relation, it can be shown that the stationary distribution follows a power law,
\begin{align}\label{solution resetting long time}
     P_r^{ss}(x|x_r) =
     \frac{r \sigma^2}{\alpha \sigma^2 + \left(\mu - \frac{\sigma^2}{2}\right)}\left\lbrace\begin{array}{l l l}
     \smallskip & \left( \frac{x}{x_r}\right)^{-\alpha-1}, \quad & x>x_r, \\ 
     & \left( \frac{x}{x_r}\right)^{\alpha+2\left(\mu - \frac{\sigma^2}{2}\right)-1}, \quad & x\leq x_r,
\end{array}\right.
     \end{align}
where
\begin{align}
    \alpha &= \frac{-(\mu - \sigma^2/2) + \sqrt{(\mu-\sigma^2/2)^2 + 2r \sigma^2}}{\sigma^2},
\end{align}
is the shape parameter. Emergence of the power law is the fingerprint of the real-world income distributions~\cite{eliazar2020power,aydiner2019money}. Other stylised facts that are recovered by the model are: larger $\mu$ (larger average population growth), larger $\sigma$ (more randomness in the dynamics) and/or smaller $r$ (less retiring or layoffs), result in a smaller  shape parameter and a heavier-tailed distribution. This leads to higher inequality and lower mobility in the economy. Thus, srGBM is a minimal model that is able  to adequately represent a range of real word situations. As such it has been implemented to date in various empirical studies~(see for example~\cite{gabaix2016dynamics,stojkoski2022income}). 

\subsection{Derivation of the first passage time in srGBM}
\label{app:srbm-mfpt-derivation}
In this section, we present the derivation for the average first passage time for the srGBM. To this end, we employ the renewal framework that allows us to write the observables in the presence of resetting in terms of the observables with $r=0$. Naturally, the bedrock of these studies relies upon obtaining exact results for the resetting free process. Since we are interested in the first passage quantities, the building blocks are usually based on the first passage time density or the survival probability function. In what follows, we first compute these quantities exactly for $r=0$ case, and then make the connection to the framework to obtain the moments.

Our starting point is to write the Fokker Planck equation for the distribution of the random income variable $x(t)$ that follows the geometric Brownian motion with $r=0$. This reads
\begin{align}
    \frac{\partial}{\partial t} P_0(x,t|x_0,t_0)=\mathcal{L}(x_0)P_0(x,t|x_0,t_0)~,
\end{align}
where $\mathcal{L}(z)=\mu z \frac{\partial}{\partial z}+\frac{\sigma^2}{2} z^2 \frac{\partial^2}{\partial z^2}$ is the generator for the GBM process (following It\^{o} convention). Computation of the first passage time density requires one to set the appropriate boundary conditions. This can be done by assuming that the GBM is constrained on some domain $\mathcal{D}$. A useful measure to compute the first passage properties is the so-called
survival probability $Q(x_0,t)$ which can be defined in one dimension (without any loss of generality) as follows -- the  probability that the income $x(t)$ has not reached a threshold $y$ up to time $t$ starting from a initial value $x_0$. Formally, this is defined as 
\begin{align}
    Q(x_0,y,t)=\int_{x_0}^y dx ~P_0(x,t|x_0,t_0)~,
    \label{surv-n}
\end{align}
so that it satisfies the Fokker-Planck equation
\begin{align}
    \frac{\partial}{\partial t} Q(x_0,y,t)=\mathcal{L}(x_0)Q(x_0,y,t)~.
    \label{survival-propagator}
\end{align}
The above equation is supplemented by the following conditions
\begin{align}
    \text{Initial condition}~~Q(x_0,y,t=0)&=1,~~x_0 \neq y, \label{initial-c} \\\text{Boundary condition}~~Q(x_0=y,y,t)&=0,~~\forall t.
    \label{boundary-c}
\end{align}
The first passage time density $f_T(t)$ is related to the survival probability via \cite{redner2001guide}
\begin{align}
f_T(t)=-\frac{\partial}{\partial t} Q(x_0,y,t)
\label{FPT-surv}
\end{align}
and its moment generating function (MGF) is given by
\begin{align}
    \tilde{T}(x_0,y,s) \equiv \langle e^{-sT} \rangle=\int_0^\infty~dt~e^{-st}~f_T(t)~.
\end{align}
The MGF is also related to the survival function in Laplace space quite trivially -- by taking the Laplace transform in Eq. (\ref{FPT-surv}), we have
\begin{align}
    \tilde{T}(x_0,y,s)=-s q(x_0,y,s)+1~,
    \label{FPT-surv-LT}
\end{align}
where $q(x_0,y,s)=\int_0^\infty dt\,e^{-st}Q(x_0,y,t)$ is the Laplace transform of the survival probability. Further, we have used the initial condition (\ref{initial-c}). The Laplace transform then trivially satisfies [see \eref{survival-propagator}]
\begin{align}
    \left[ \mathcal{L}-s   \right]q(x_0,y,s)=-1~,
\end{align}
with appropriate boundary conditions. Translating to the first passage function, we arrive at the following eigenvalue equation for the moment generating function
\begin{align}
    \mathcal{L} \tilde{T}(x_0,y,s)=s\tilde{T}(x_0,y,s)~,
    \label{FP-eqn}
\end{align}
We now assume that in one dimension, the threshold is at $y>0$ so that 
\begin{align}
    \tilde{T}(x_0=y,y,s)=1.
\label{BC-1}
\end{align} 
Moreover, near $|x_0| \to 0$, we should have
\begin{align}
    \tilde{T}(x_0,y,s)< \infty~
    \label{BC-2}
\end{align}
since starting from $x_0 \to 0^+$, it will take a \textit{finite time} for the stochastic process to hit $y$, and thus the first passage time can not diverge. The solution for \eref{FP-eqn} with the above boundary conditions reads
\begin{align}
    \tilde{T}(x_0,y,s)&= A x_0^{q_1(s)}+B x_0^{q_2(s)},\\
    \text{where} \quad
    q_1(s)&=\frac{\sqrt{(\sigma^2-2\mu)^2+8s \sigma^2}+(\sigma^2-2\mu)}{2 \sigma^2}>0 \\
    \text{and} \quad
    q_2(s) &=-\frac{\sqrt{(\sigma^2-2\mu)^2+8s \sigma^2}-(\sigma^2-2\mu)}{2 \sigma^2}<0,
\end{align}
for any combination of $\mu$ and $\sigma^2$.  
To be consistent with the boundary condition (\ref{BC-2}), we should have $B=0$, and thus
\begin{align}
    \tilde{T}(x_0,y,s)=\langle e^{-sT} \rangle=A x_0^{q_1(s)},
\end{align}
where $A$ can be calculated using the other boundary condition (\ref{BC-1}). Putting all the pieces together, we have
\begin{align}
   \tilde{ T}(x_0,y,s)=\left( \frac{x_0}{y} \right)^{q_1(s)}, \quad x_0 \leq y.
   \label{MGF-LT}
\end{align}
\eref{MGF-LT} is a very useful result since it allows us to compute all the moments for the first passage times by noting 
\begin{align}
\langle T^n(x_0,y) \rangle \equiv (-1)^n \frac{d}{ds}  \tilde{ T}(x_0,y,s)|_{s \to 0},
\label{moments}
\end{align}
which will also be useful to obtain the first passage time moments in the presence of resetting (see below).

The first passage quantities under resetting (i.e., $r>0$) can be related to the underlying reset-free (i.e., $r=0$) quantities using the renewal formalism \cite{pal2017first,pal2016diffusion,evans2020stochastic,reuveni2016optimal,bonomo2022mitigating}. 
Our starting point is to write a renewal equation for the survival probability $Q_r(x_0,y,x_r,t)$ (similar to $Q(x_0,y,t)$ in Eq. (\ref{surv-n})) which is defined as the probability that the process has started from $x_0$, got reset at $x_r$ and stayed below the boundary coordinate $y$ upto time $t$. The renewal equation for the survival function reads \cite{pal2016diffusion}
\begin{align}
    Q_r(x_0,y,x_r,t)=e^{-rt} Q(x_0,y,t)+\int_0^t~d\tau~re^{-r\tau}~Q(x_0,y,\tau)~ Q_r(x_r,y,x_r,t-\tau).
    \label{ND-1}
\end{align}
Taking Laplace transform on the both sides of Eq. (\ref{ND-1}) and after some manipulations we arrive at the following relation
\begin{align}
    q_r(x_0,y,x_r,s)&=\frac{q(x_0,y,s+r)}{1-rq(x_r,y,s+r)} \\
    &=\frac{1-\tilde{T}(x_0,y,s+r)}{s+r \tilde{T}(x_r,y,s+r)},
\end{align}
where, while going from the first line to second, we have used Eq. (\ref{FPT-surv-LT}) i.e., the relation between first passage time density and survival function in the Laplace space. The mean first passage time is then given by
\begin{align}
\langle T_r (x_0,y,x_r)\rangle = \int_0^\infty~dt~tf_{T_r}(t)=q_r(x_0,y,x_r,s \to 0)= \frac{1-\tilde{T}(x_0,y,r)}{r \tilde{T}(x_r,y,r)},
\label{Mean-Poiss-diff}
\end{align}
where $\tilde{ T}(x_0,y,r)=\left( \frac{x_0}{y} \right)^{q_1(r)}$. This is the central formula behind the MFPT analysis that was announced in the main text. We have verified \eref{Mean-Poiss-diff} with numerical simulations in Fig. (\ref{fig:mfpt-theoretical}). The results show an excellent agreement between them.

A simple observation at the Fig. (\ref{fig:mfpt-theoretical}) shows us that the MFPT can be optimized as a function of the resetting rate $r$. This optimal rate can be computed from the relation 
\begin{align}
    \frac{\partial}{\partial r} \langle T_r(x_0,y,x_r) \rangle \bigg|_{r=r^*}=0.
\end{align}
Although it is not possible to obtain an exact expression for $r^*$, one can use a numerical approach. The optimal resetting shows interesting behaviors in terms of the intrinsic system parameters such as $\mu$ and $\sigma$ as can be seen from Fig. \ref{fig:mfpt-theoretical}. While the optimal rate is non-zero for small values of $\mu$, it eventually becomes zero for large $\mu$. On the other hand, this behavior is reversed as $\sigma$ is varied from small to large. Such a transition of the optimal resetting rate from a finite value to zero is reminiscent of the canonical thermodynamic phase transition that can further be understood within the Landau like mean field theory (see \cite{pal2019landau} for a general theory and such instances in physics).

\subsection{Numerical Estimation of MFPT through srGBM}
\label{app:mfpt-srGBM}

The basic ingredient used to numerically simulate srGBM is to generate a trajectory using Eq.~\eqref{eq:srgbm-microscopic}. Concretely, to obtain the distribution of the position of the particle at time $t$, we discretize the time $t = n\Delta t$, where $n$ is an integer. We initialize the position of the particle at $x(0)=1$, and then, at each step $(\tau = 1,...,n)$, the particle can either reset or it can evolve according to the laws of GBM. Thus,

\begin{enumerate}
    \item with probability $1-r\Delta t$ ($r$ is the rate of resetting); the particle undergoes GBM so that
\begin{align}\label{GBM1}
    x(\tau\Delta t) = x[(\tau-1)\Delta t][\mu+\sigma \sqrt{\Delta t}\eta(\tau\Delta t)],
\end{align}
    where $\eta(\tau \Delta t)$ is a Gaussian random variable with mean 0 and variance 1, and $\Delta t$ is the microscopic time step;
    \item with complementary probability $r\Delta t$, resetting occurs such that
\begin{align}\label{GBM2}
    x(\tau\Delta t) = x(0) = 1.
\end{align}
\end{enumerate}

\begin{figure}[H]
\centering
\includegraphics[width=10cm]{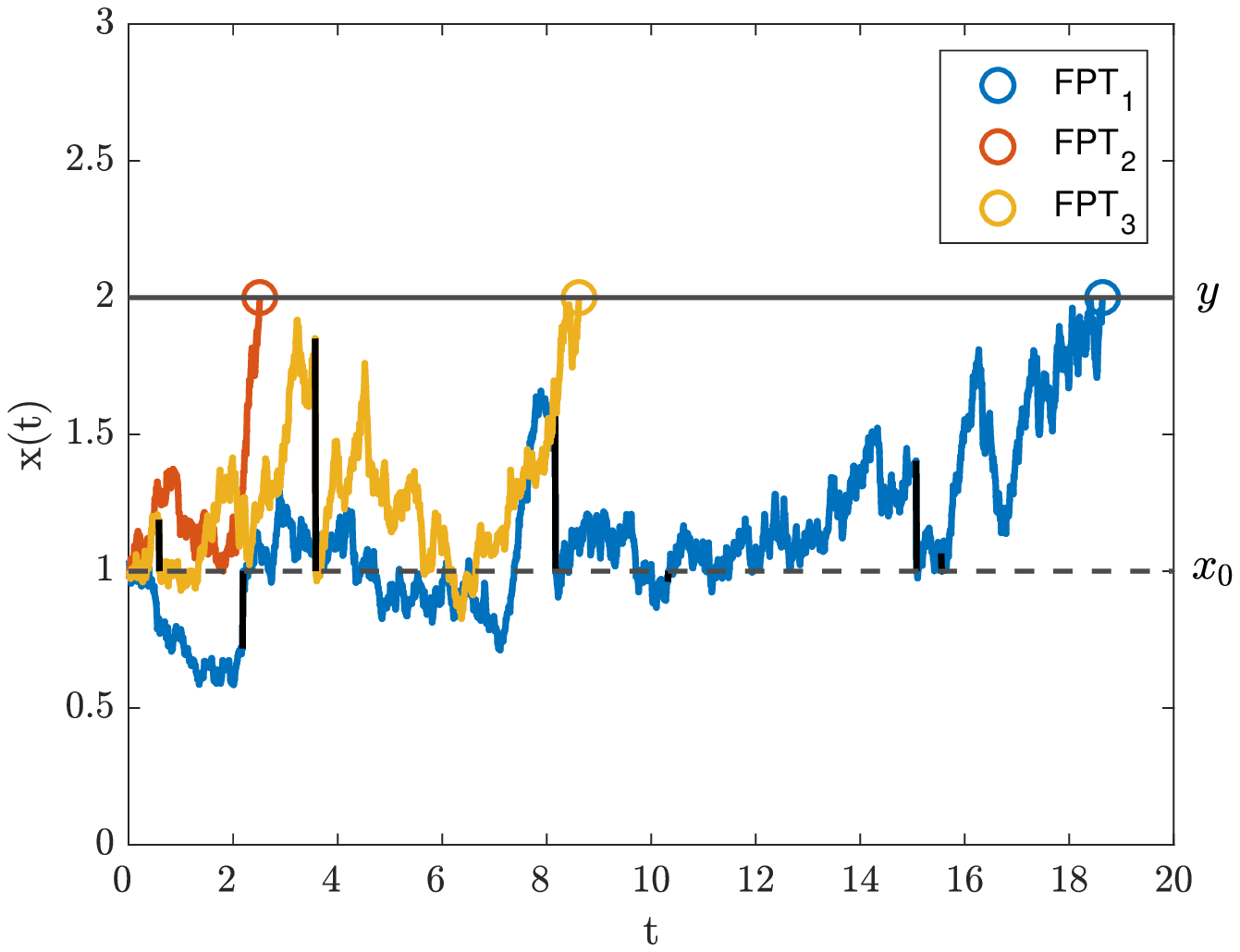}
\caption{\textbf{First passage time events.} Here we plot three typical srGBM trajectories with their corresponding first passage times (circled points). Every trajectory starts at $x_0=1$ (horizontal dashed line). The horizontal bold line at $x=2$ represents the predefined target. The black vertical lines represent a resetting event. Here, we have assumed $x_0=x_r$ without any loss of generality.}\label{fig:fpt-events}
\end{figure}

The length of a single first passage event can be estimated as follows: we simulate the stochastic process described in Eqs.~\eqref{GBM1} and \eqref{GBM2} until the particle hits a predefined virtual absorbing boundary, $x(\tau\Delta t)=y$. The length of this first passage event is simply $\tau\Delta t$. In order to illustrate how we can numerically extract a series of first passage time events, we identify the sections of three sample trajectories which start at $x(0)=1$ and end at a predetermined boundary, $x(\tau\Delta t)=y$, that is represented by the solid line in Fig.~\ref{fig:fpt-events}. Given these first passage events, we can easily calculate the corresponding first passage times by taking their time span. The average of these time lengths gives an estimate for the MFPT of the stochastic process and generally can be written as:
\begin{align}
    \langle\hat{\tilde{T_r}}\rangle = \frac{\Sigma_{N}\, \tau\Delta t}{N}.
\end{align}
As we increase the number of sampled trajectories, $N$, the estimated empirical MFPT should converge to the analytical value.

\subsection{Estimation of the MFPT through transition matrices}
\label{app:mfpt-matrices}

An income transition matrix aggregates the income rankings and summarizes mobility in a
stochastic matrix $\mathbf{A}$ in which the elements $A_{kl}$ quantify the probability that an individual in income quantile $k$ in period $t$ is found in income quantile $l$ in period $t + \Delta$.

\begin{figure}[H]
\centering
\includegraphics[width=8.7cm]{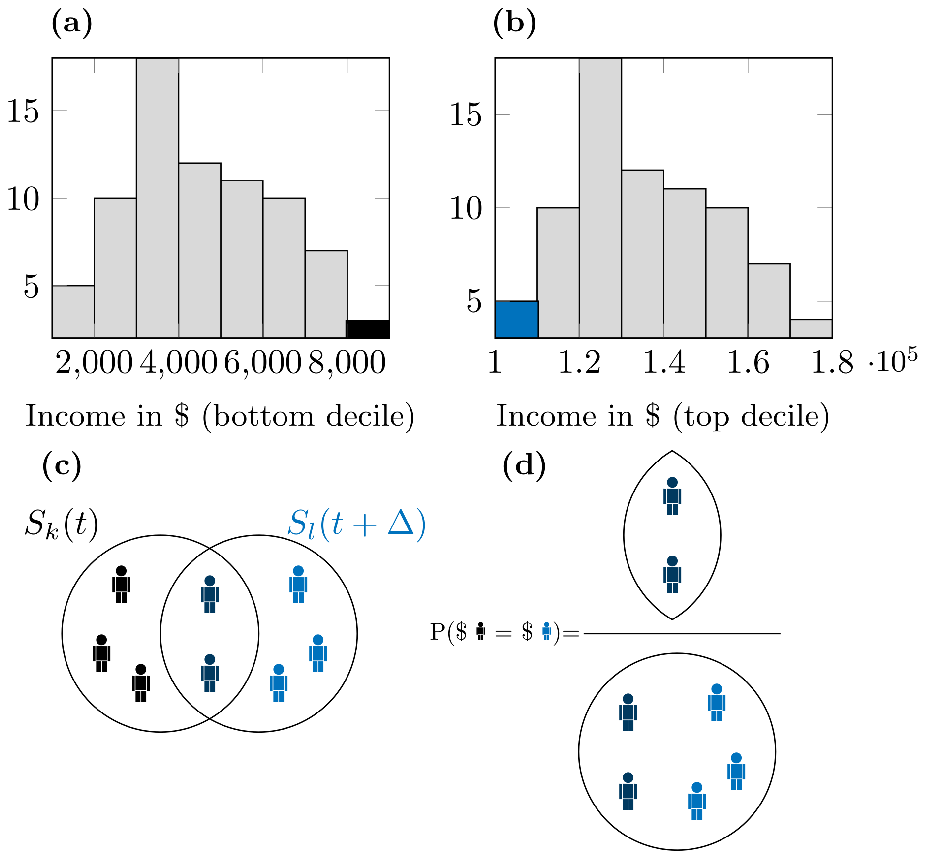}
\caption{\textbf{Construction of a transition matrix.} \textbf{(a)} Income distribution histogram for the lowest paid individuals at time $t$. \textbf{(b)} Income distribution histogram for the highest paid individuals at time $t+\Delta$. \textbf{(c)} Left diagram: the set of individuals that belong to the bottom quantile at $t$. Right diagram: the set of individuals that belong to the top quantile at $t+\Delta$. \textbf{(d)} Transition probability from the bottom to the top quantile in time $\Delta t$ is calculated as the fraction of individuals that reached the threshold income of the target quantile, namely from the black colored bar in \textbf{(a)} to the blue colored bar in \textbf{(b)}.}
\label{fig:transition_matrix}
\end{figure}

Mathematically, the entries of the transition matrix can be defined as follows. Let $\mathcal{S}_k(t)$ denote the set of individuals that are part of quantile $k$ in time period $t$. Then,
\begin{align}
    A_{kl} = \frac{|\mathcal{S}_k(t) \cap \mathcal{S}_l(t+\Delta) | }{|\mathcal{S}_l(t+\Delta) |},
\end{align}
where $|\mathcal{S}|$ is the cardinality of $\mathcal{S}$. For example. suppose that we want to calculate the element $A_{1,10}$ of the transition matrix. Visually, this entry represents the probability that the highest paid worker of the bottom quantile (see Fig.~\ref{fig:transition_matrix}\textbf{(a)}) in time $t$ reaches the threshold income of the lowest paid worker of the top quantile (see Fig.~\ref{fig:transition_matrix}\textbf{(b)}) in time $t+\Delta t$. This probability is calculated as a fraction of the individuals that cross the minimum income of the target decile in $\Delta t$ as shown in Fig.~\ref{fig:transition_matrix}\textbf{(c)} and \textbf{(d)}. We can easily generalise this procedure and compute the full transition matrix $\textbf{A}$ after arbitrary time steps.

The income transition matrix that is generated from real world data follows approximately continuous time dynamis. The standard methods for estimating the MFPT from a transition matrix, however, assume discrete time.
Conveniently, $\mathbf{A}$ can be easily transformed to a so called embedded Markov chain matrix $\mathbf{\tilde{A}}$ whose MFPT estimated using discrete time methods is the same as if we were to estimate the continuous time MFPT from $\mathbf{A}$.

In what follows, we describe the procedure for generating $\mathbf{\tilde{A}}$ and estimating the corresponding MFPTs.

The fist step is to generate the Markov generator matrix $\mathbf{Q}$ of $\mathbf{A}$ that describes the rates at which workers move across quantiles. The elements $Q_{kl}$ of $\mathbf{Q}$ are
  \[
   Q_{kl} = \Bigg\{\begin{array}{lr}
        \log(A_{kl}), & \text{if } k=l\\
        A_{kl}\log(A_{kl})/(1 - A_{kl}), & \text{otherwise}.
        \end{array}
  \]
Using $\mathbf{Q}$, $\mathbf{\tilde{A}}$  can be estimated as
\begin{align}
    \mathbf{\tilde{A}} &= \mathbf{I} - \texttt{diag}\left( \mathbf{Q} \right)^{-1} \mathbf{Q},
\end{align}
where $\mathbf{I}$ is the identity matrix and $\texttt{diag}\left( \mathbf{Q} \right)$ is a matrix with entries equal to $Q_{kl}$ if $k = l$, and $0$ otherwise.

The MFPT between two quantiles $k$ and $l$ of a transition matrix $A$ with $K$ quantiles is related to the vector that describes the stationary transition probabilities $\mathbf{\pi}^{'}=(\pi_1,\pi_2,...,p_K)$ of the discrete Markov chain given with $\mathbf{\tilde{A}}$.  The matrix $\mathbf{M}$ whose entries $M_{kl}$ give the MFPT between $k$ and $l$ can be estimated in the following way:
\begin{enumerate}
    \item Compute the stationary probability vector $p^{*t}$ of the Markov chain described with $\mathbf{\tilde{A}}$. This vector is given by the left eigenvector corresponding to the largest eigenvalue of $\mathbf{\tilde{A}}$.
    \item Compute the fundamental matrix $\mathbf{Z}$, as $\mathbf{Z}=[\mathbf{I}-\mathbf{A}+\mathbf{1} \mathbf{\pi}^{'}]^{-1}$.
    \item Compute the MFPT matrix $\mathbf{M}$ with mean first passage time from quantile $k$ to quantile $l$ defined as $M_{k,l} = \frac{Z_{ll} - Z_{kl}}{\pi_l}$. 
\end{enumerate}

We refer the reader to Refs.~\cite{israel2001finding,hunter2018computation,fronczak2009biased,noh2004random} for more details about the estimation procedure. 




\subsection{Method for empirical estimation of srGBM parameters}\label{app:srgbm-empirical-methodology}

We assume that the income dynamics follows srGBM that is constantly under the threat of changing its parameters. In this context, we will assume that the resetting rate $r(t)$ is a function of time and will provide an approximation $\hat{r}(\tau)$ with the fraction of people that lost and/or left their job. For simplicity, we will measure the resetting rate on a yearly basis and assume that in between two years the resetting rate is fixed, i.e., $r(t) \approx \hat{r}(\tau)$ for any $\tau$ between $t$ and $t+1$. Our goal is to simultaneously provide consistent estimates $\hat{\mu}(\tau)$ and $\hat{\sigma}(\tau)$ for the drift parameter $\mu(\tau)$ and the noise amplitude $\sigma(\tau)$ as a function of the time, that best fits the observed shares of income owned by the top 1\% in the US income distribution. The assumption for dynamics in the model parameters reflects the possibility of noise in the data. In addition, it can be an approximation for the changes in economic conditions that affect the srGBM dynamics. These can be either due to changes in government policies or due to circumstances that are not under the control of the policymakers.

Formally, the estimation procedure consists of the following steps: 

\textbf{Step 1}: Fix the resetting rate $\hat{r}(0)$ in the initial period at the initial year $\tau=0$ and then estimate $\hat{\mu}(0)$ and $\hat{\sigma}(0)$ to match the srGBM stationary distribution.

\textbf{Step 2}: Propagate $N$ individual income trajectories according to the laws of srGBM. That is, with probability $1-\hat{r}(\tau)\Delta t$ the income undergoes GBM so that: 
\begin{align}\label{simulations}
x_i(t+\Delta t) = x_i(t) + x_i(t)[\hat{\mu}(\tau) \Delta t + \hat{\sigma}(\tau)\sqrt{\Delta t}\eta_i(\Delta t)],
\end{align} 
where $\eta_i(\Delta t)$ is a Gaussian random variable with zero mean and unit variance, and $\Delta t$ is a small time increment. Here, we used the It\^{o} convention. With complementary probability $\hat{r}(\tau)\Delta t$, the income resets to the initial position:
\begin{align}\label{simulations2}
x_i(t+\Delta t) = 1.
\end{align}
At last, we find the values $\hat{\mu}(\tau+1)$ and $\hat{\sigma}(\tau+1)$ that minimise the squared difference between the inferred share of the top 1\% in the modelled population in year $\tau+1$ and the observed share in real data.

\textbf{Step 3}: Repeat Step 2 until the end of the time series.

For each time series we run a simulation for a model economy of $N=10^6$ workers. However, because of the randomness of the numeric simulations, each simulation will result in different fitted values. To take this into consideration, we construct a Monte Carlo estimation by repeating the process 100 times and report the average value of $\hat{\mu}(\tau)$ and $\hat{\sigma}(\tau)$. In addition, this allows us to estimate the variability of the results and provide confidence intervals for both parameters.

\subsection{Method for empirical estimation of srGBM MFPT}
\label{app:empirical-srgbm-MFPT}

The estimation procedure for the srGBM MFPT consists of the following steps: 
\begin{enumerate}
\item Fix the srGBM parameters to match those estimated in \cite{stojkoski2022income}. Choose an initial position and a target. These positions represent the estimated income for the starting and target percentile, respectively.
\item Calculate the MFPT using Eq.~\eqref{eq:srgbm-mfpt}.
\item Repeat Step 2 until the end of the time series.
\end{enumerate}

\subsection{Method for estimation of Fraction of individuals reaching a target income in X years}\label{app:srgbm-empirical-fraction}

The estimation procedure for the fraction of individuals that reach a target income in X consists of the following steps:
\begin{enumerate}
\item Fix the srGBM parameters to match those estimated in \cite{stojkoski2022income}. Choose an initial position and a target. These positions represent the average income for the poorest individuals and target decile, respectively.
\item Propagate $N$ individual income trajectories according to the laws of srGBM for XX years. Calculate how many individuals $n$ out of the population $N$ reach the target income fixed in the first step.
\item Repeat Step 2 until the end of the time series.
\end{enumerate}

\subsection{Method for estimation of the optimal resetting rate} \label{app:srgbm-empirical-optimal-rate}

In order to estimate the optimal resetting rate for each year, we have to fix $\hat{\mu}(\tau)$, $\hat{\sigma}(\tau)$, $x_r(\tau)$, $x_0(\tau)$, and the target $y(\tau)$. However, given that there are multiple choices for a starting and target position, we can estimate in total 45 different optimal resetting rates. The minimization procedure is performed using Hessian-free optimization, in particular the truncated Newton (TNC) algorithm.

\end{document}